\documentclass{elsarticle}

\usepackage{lineno,hyperref}
\modulolinenumbers[5]

\journal{journal}

\usepackage{amsmath}
\usepackage{subcaption}
\usepackage{amssymb}
\usepackage{wasysym}
\usepackage{xcolor}
\definecolor{myred}{rgb}{1,0,0}
\definecolor{myorange}{HTML}{ff7f0e}
\definecolor{mygreen}{HTML}{2ca02c}
\newcommand{\rev}[1]{#1}
\newcommand{\Diff}[2]{\frac{\mathrm{D} #1}{\mathrm{D} #2}}
\newcommand{\diff}[2]{\frac{\partial #1}{\partial #2}}

\newcommand{\vect}[1]{\boldsymbol{#1}}
\newcommand\Tstrut{\rule{0pt}{2.6ex}}         
\newcommand\Bstrut{\rule[-0.9ex]{0pt}{0pt}}   

\makeatletter
\def\ps@pprintTitle{%
	\let\@oddhead\@empty
	\let\@evenhead\@empty
	\def\@oddfoot{}%
	\let\@evenfoot\@oddfoot}
\makeatother

\usepackage{tikz}
\newcommand{\sqdiamond}[1][fill=myorange]{\tikz [x=1.2ex,y=1.2ex,line width=.1ex,line join=round, yshift=-0.285ex] \draw  [#1]  (0,.5) -- (.5,1) -- (1,.5) -- (.5,0) -- (0,.5) -- cycle;}%
\newcommand{\tikzcircle}[2][red,fill=red]{\tikz[baseline=-0.5ex]\draw[#1,radius=#2] (0,0) circle ;}
\begin{document}

\begin{frontmatter}
		
	  \title{High-order overset grid method for detecting particle impaction on a cylinder in a cross flow}
		
  \author[mymainaddress]{J\o rgen R.~Aarnes}
  \cortext[mycorrespondingauthor]{Corresponding author: J\o rgen R\o ysland Aarnes}
  \ead{jorgen.r.aarnes@ntnu.no}
  
  \author[mymainaddress,mysecondaryaddress]{Nils E. L. Haugen}
  \author[mymainaddress]{Helge I. Andersson}

  \address[mymainaddress]{Department of Energy and Process Engineering, Norwegian University of Science and Technology, 7034 Trondheim, Norway}
  \address[mysecondaryaddress]{SINTEF Energy Research, 7465 Trondheim Norway}
  
  \begin{abstract}
    An overset grid method \rev{was} used to investigate the interaction between a particle-laden flow and a circular cylinder. The overset grid method is implemented in the Pencil Code, a high-order finite-difference code for compressible flow simulation. High-order summation-by-part operators \rev{were} used at the cylinder boundary, and both bi-linear Lagrangian and bi-quadratic spline interpolation \rev{was} used \rev{to} communicate between the background grid and the body-conformal cylindrical grid. The performance of the overset grid method \rev{was} assessed \rev{to} benchmark cases of steady and unsteady flows past a cylinder. For steady flow at low Reynolds number, high-order accuracy \rev{was} achieved for velocity components. Results for flow in the vortex shedding regime showed good agreement \rev{to the literature}. The method \rev{was also} applied to particle-laden flow simulations, where spherical point particles \rev{were} inserted upstream of the cylinder. These inertial particles \rev{were} convected towards and (possibly) past the cylinder. The simulations reproduced data from \rev{the} literature at a significantly reduced cost, revealing that the previously published DNS data is less accurate than assumed for particles with very small Stokes numbers.
  \end{abstract}
		
		
\end{frontmatter}
	
\section{Introduction}
A common flow problem in numerical simulations is flow past a bluff body. Obstructions in the flow include (but are not limited to) spheres, flat plates, circular, rectangular or elliptical cylinders, triangles, spheroids, and complex geometries made out of a combination of these \rev{shapes}.  Particle-laden flows interacting with such obstacles are important for a range of applications. Whether the goal is to maximize the particle extraction from the flow, as for filter applications, or \rev{to minimize} particle \rev{attachment} on the object to avoid an insulating layer, as for biomass boilers, understanding the mechanics of inertial particles helps improve design, and hence, \rev{the} efficiency of said applications.  Accurate prediction of particle behavior in the vicinity of bluff bodies require\rev{s} highly accurate boundary layer representation \rev{with}in numerical simulations. Finding the numerical method best suited to this task is not trivial, \rev{and} can have a huge impact on both \rev{the} efficiency and accuracy of simulations.

\subsection{Representing solid objects in the flow}
For generic shapes (cylinders, spheres, plates, etc.) body-fitted structured meshes are commonly used to accurately resolve the solid boundary. \rev{These} methods use grids that conform to the solid (or solids) immersed in the flow and to other physical boundaries of the domain (inlet/outlet, walls, etc.). Depending on the domain \rev{geometry}, this may require some deformation of the grid to conform to the boundaries, in addition to the mapping of the flow domain onto a simple computational domain. The result may \rev{lead to} unnecessary local variations \rev{in} the grid and \rev{rather} time consuming grid generation \cite{versteeg2007introduction}. Alternatively, unstructured meshes \rev{can be applied} to resolve the solid boundaries in the flow. Unstructured meshe\rev{s} provide the highest flexibility in adapting a mesh to the flow problem, and are a good alternative for complex geometries when finite-volume or finite-element formulations of the governing equations are used \cite{mavriplis1997unstructured}. Among the disadvantages of such grids are much larger storage requirements \cite{pletcher2012computational} and the need for intricate mesh generation techniques \cite{owen1998survey}. 

An alternative to body-fitted grids are non-conforming (typically Cartesian) meshes, where a solid in the flow is represented by \rev{a} change in the fluid equations in the vicinity of the solid boundary. One such method, \rev{which} has gained vast popularity \rev{over} the last decades, is the immersed boundary method (IBM). This method (or rather, this class of methods) was originally developed to model flow around heart valves \cite{Peskin1972} by allowing for \rev{the} representation of bluff bodies in the flow without using \rev{a} body-conformal grid. A simple Cartesian grid can be used, where the boundary conditions (the sharp interface) of the bluff body are incorporated in\rev{to} the solver by a modification of the equations in the vicinity of the boundary (see review article by Mittal \& Iaccarino \cite{Mittal2005} and references therein for details). This makes IBMs very flexible \rev{for} representing bluff bodies, and particularly well-suited \rev{to} complex geometries, where the use of body-fitted structured meshes \rev{is} limited. A caveat \rev{to} the IBM is the difficulty in achieving high-order accuracy near boundaries that do not conform to physical boundaries. For complex geometries this may be regarded as a necessary loss in order to be able to represent the boundary. For flow past \rev{simple geometric,} bodies other methods may be more suitable, especially when the accuracy in the vicinity of the surface is a major concern.
	
Roughly ten years after the development of the IBM, a method of multiple grids \textit{overset} \rev{on top of} one another was proposed to represent solids in a flow (see \cite{Steger1983,Benek1985,Steger1987}). Overset grids, or Chimera methods, employ body-conformal grids at the bluff bodies, but the body-conformal grids do not extend to the domain boundaries. Instead, a non-conforming background grid (typically uniform Cartesian) is used, and updated flow information \rev{within} overlapping grid regions is communicated between grids at every time step. In this way, \rev{the} flow simulation is split into multiple \rev{sub-}simulations, one for each grid, and the boundaries of one grid \rev{are} updated with information from the other grids.  The background grid is used to compute the general flow \rev{field} outside the smaller body-fitted grids, and the communication between grids is done \rev{through} interpolation.
	
Overset grid methods have the advantage of being highly accurate at the solid-fluid interface. This is due to the use of body-fitted grids in these regions, and the flexibility in grid stretching made possible when several grids are used. At the same time, no grid deformation is necessary to conform to domain boundaries, due to the use of an appropriate \rev{non-conformal} background grid. If the domain is circular, a cylindrical grid can be used as a background grid, if rectangular, a Cartesian grid, etc. 

The communication between the grids is the limiting factor in terms of \rev{the} accuracy of overset grid methods. \rev{In general, the} interpolation of flow variables is detrimental to mass conservation (although conservative, mass correcting overset grid methods do exist for finite-volume codes, see e.g. \cite{Part-Enander1994,Zang1995}). Using high-order interpolation between grids have proved beneficial in regards to the overall accuracy and stability of the overset grid method for both finite-difference and finite-volume implementations \cite{Sherer2005,Chicheportiche2012,Volkner2017}. While advantageous in terms of accuracy, high-order interpolation techniques have the disadvantage of increase\rev{s} in complexity, inter-processor communication and floating-point operations, \rev{when} compared to low-order interpolation schemes. Furthermore, straightforward extension to high-order interpolation, typically from second-order to fourth-order Lagrangian interpolation, does not guarantee a better solution. Possible overshoots in the interpolation polynomials may have a devastating impact on the interpolation accuracy. The applied interpolation scheme should therefore be evaluated for the \rev{specific} flow problem at hand. For overset grid implementations, several interpolation schemes are available. In this study \rev{two such schemes are compared}: bi-linear Lagrangian interpolation and bi-quadratic spline interpolation. Together with high-order low-pass filtering, the resulting computations \rev{were} both stable and accurate. This topic \rev{is} further \rev{discussed} in Section \ref{sec:method}.

If several body-fitted grids overlap, the overset grid computations \rev{become} increasingly difficult, particularly in regards to the communication \rev{between} the different grids. For \rev{the purposes of this paper}, the discussion \rev{is limited} to a single body-fitted grid on top of a Cartesian background grid. For more general discussion on overset grids, see \cite{Meakin1995} or \cite{Chesshire1990}.

\subsection{Particle impaction}
When considering particle deposition on a surface, two mechanisms are \rev{required} for a particle to deposit. The particle must \rev{first} impact the surface, that is, it must physical\rev{ly} contact the surface, and then it must \rev{adhere} to \rev{the surface}. Only the first of these two mechanisms will be \rev{the} focus of this study. Hence, all particles that come in\rev{to} contact with the bluff body are \rev{considered to have been} absorbed by it. Further, only inertial impaction is considered. \rev{Any} other particle impaction mechanisms \rev{including} Brownian motion, thermophoresis and turbulent diffusion are omitted. \rev{Note that this is not an acceptable omission in non-isothermal flows, where the effects of temperature will be large on small particles} (see \cite{GarciaPerez2016,beckmann2016measurements}).
	
The impaction efficiency $ \eta = N_{imp}/N_{ins}$ is a measure of the cylindrical object\rev{'s ability} to capture the particles that \rev{are} initially incident on the cylinder. The number of impacting particles is given by $N_{imp}$, while $ N_{ins} $ is \rev{the} count of particles with a center of mass that is initially moving in the direction of the solid object. Note that following this convention may lead to $ \eta > 1 $, even if no forces act on the particles, since a particle may follow a path close enough to be intercepted by the object, due to its finite size, \rev{even though} the center of mass does not hit the object.
	
A \rev{fluid} flow will be deflected by the object, and particles in the flow will experience a drag force. This force will accelerate the particles \rev{along} the fluid trajectory, leading particles away from the bluff body. The particle Stokes number, $ St = \tau_p/\tau_f $, where $ \tau_p $ and $ \tau_f $ are particle and fluid time scales, respectively (details in Section \ref{sec:particles}), can be considered a measure of particle inertia. Hence, particles with a small Stokes number follow the flow to a larger extent than particles with a large Stokes number.  By using potential flow theory to compute the flow past a circular cylinder, Israel \& Rosner \cite{Israel2007} \rev{determined a relation} for the impaction efficiency as a function of the Stokes number. The predictions by Israel \& Rosner are inaccurate in predicting particle impactions for flows where the viscous boundary layer \rev{of} the cylinder plays an \rev{significant} role. \rev{This is because } potential flow theory assumes inviscid flow. In particular, th\rev{is} theory is insufficient \rev{at} predicting particle impactions for particles with small Stokes number\rev{s}, and \rev{for} moderate Reynolds number flows. Here, the Reynolds number is defined as $ Re=U_0 D/\nu $, where $ U_0 $ is the mean flow velocity, $ D $ is the diameter of a cylinder (the bluff body in the flow) and $ \nu $ is the kinematic viscosity. Haugen \& Kragset \cite{Haugen2010} performed simulations using the Pencil Code to compute inertial particle impaction on a cylinder in a crossflow for different Stokes and Reynolds numbers. Later, Haugen et al. \cite{Haugen2013} performed a similar study on a flow with multiple cylinders\rev{,} in order to emulate impaction on a super-heater tube bundle. The impaction efficiencies obtained by Haugen \& Kragset \cite{Haugen2010} have been used as benchmarking results, but \rev{were} limited to moderate Reynolds numbers and two-dimensional flows. Part of the reason \rev{for} this limitation is the use of an immersed boundary method that requires a very fine grid to achieve the \rev{required} accuracy.
	
\subsection{Present}
The purpose of this paper is to introduce an overset grid method applicable to compressible particle-laden flows past a circular cylinder, and to assess its performance in benchmarking cases and \rev{a true} particle-laden flow simulation. The method has been implemented in the open source compressible flow solver known as the Pencil Code \cite{Brandenburg2002,pencilcode}\rev{, with the} aim to improve the accuracy in the vicinity of \rev{the} cylinder and to reduce \rev{the} computational cost of particle-laden flow simulations.
	
The structure of the paper \rev{as} follows: In Section \ref{sec:method} the equations governing the flow and the bluff body representation \rev{are} described. An assessment of the accuracy of the method for steady and unsteady flow past a cylinder is given in Section \ref{sec:benchmarking}. In Section \ref{sec:particles} the capabilities of the overset grid method \rev{are} demonstrated by simulating particle-laden flow\rev{s} interacting with a bluff body at a moderate Reynolds number. The results and the computational costs are compared with those of Haugen \& Kragset~\cite{Haugen2010}, before concluding remarks are given in Section \ref{sec:conclusion}.
	
\section{Methodology}
\label{sec:method}
\subsection{Governing equations}
The governing equations of the flow are the continuity equation:
	
\begin{equation}
  \Diff{\rho}{t} = - \rho \vect{\nabla}\cdot\vect{u}\, , 
\end{equation}
and the momentum equation:
\begin{equation}
  \rho \Diff{\vect{u}}{t} = -\vect{\nabla} p + \vect{\nabla} \cdot \left( 2 \mu \vect{S} \right) \, , 
\end{equation}
where $ \rho $, $ t $, $ \vect{u} $ and $ p $ are the density, time, velocity vector and pressure, respectively, and $ \mu = \rho \nu $ is the dynamic viscosity. The compressible strain rate tensor $ \vect{S} $ is given by:
\begin{equation}
  \vect{S} = \frac{1}{2}\left(\vect{\nabla}\vect{u} + \left(\vect{\nabla}\vect{u}\right)^T  \right) - \vect{I}\left(\frac{1}{3} \vect{\nabla}\cdot\vect{u}\right) \, ,
\end{equation}
where $ \vect{I} $ is the identity matrix. The pressure is computed \rev{using} the isothermal ideal gas law, $ p = c_s^2 \rho \, ,$ where $ c_s $ is the speed of sound. The flow is isothermal and weakly compressible, with \rev{a} Mach number of $\sim 0.1$ for all simulations. With a constant speed of sound (for the isothermal case) and a constant kinematic viscosity, the momentum equation \rev{to be} solved on the overset grids is:
\begin{equation}
  \Diff{\vect{u}}{t} = -c_s^2 \vect{\nabla} \left(\ln \rho \right) + \nu \left(\vect{\nabla}^2 \vect{u} + \frac{1}{3}\vect{\nabla}\left(\vect{\nabla} \cdot \vect{u} \right) + 2\vect{S} \cdot \vect{\nabla} \left(\ln \rho \right) \right) \,. 
\end{equation}		
The governing equations \rev{were} discretized with sixth-order finite-differences in space and a third-order memory efficient Runge-Kutta scheme in time \cite{williamson1980low}. The flow \rev{was} simulated on a rectangular domain with \rev{an} inlet at the bottom and flow in the vertical direction. The circular cylinder \rev{was} situated in the center of the domain, \rev{with the following} boundary conditions\rev{:} no-slip and impermeability for velocity, and zero gradient in the radial direction for the density. The latter condition \rev{was} derived from the ideal gas law and the boundary layer approximation $ \left( \diff{p}{n}  = 0\, , \text{where $n$ is the wall normal direction}\right) $ for an isothermal flow. Navier-Stokes characteristic boundary conditions \rev{were} used both at the inlet and at the outlet of the flow domain. This boundary condition is a formulation that makes use of one-dimensional characteristic wave relations to allow acoustic waves to pass through the boundaries \cite{Poinsot1992,yoo2005characteristic}. The remaining domain boundaries \rev{were} periodic.

\subsection{Overset grids {\label{subsec:ogrid}}}
	
\begin{figure}[t!]
  \centering
  \begin{subfigure}[a]{\textwidth}
  	\centering
  	\includegraphics[width=0.99\linewidth]{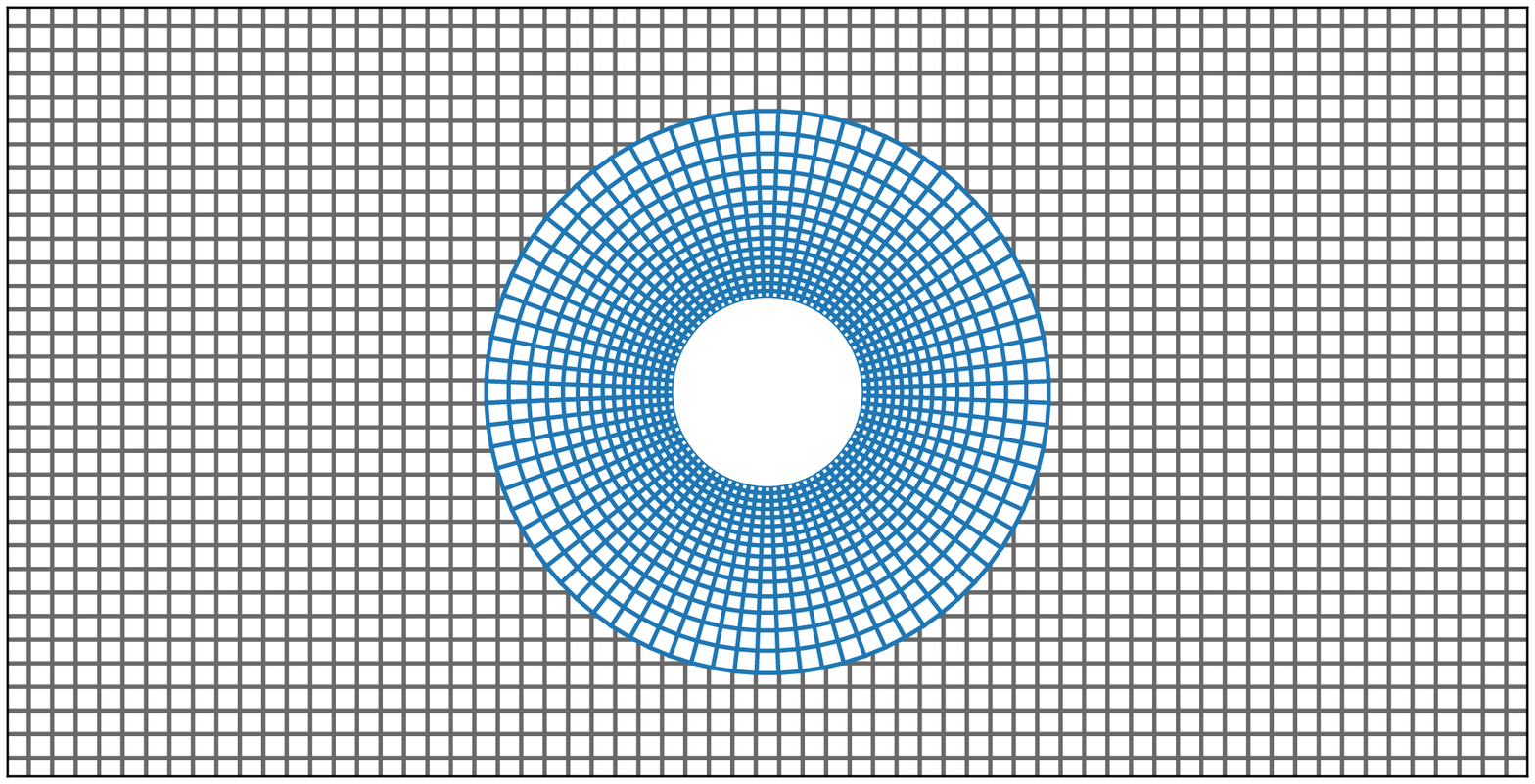}
  	\caption{}
  	\label{subfig:ogrid}
  \end{subfigure}
  
  \begin{subfigure}[a]{0.49\textwidth}
    \centering
    \includegraphics[width=0.99\linewidth]{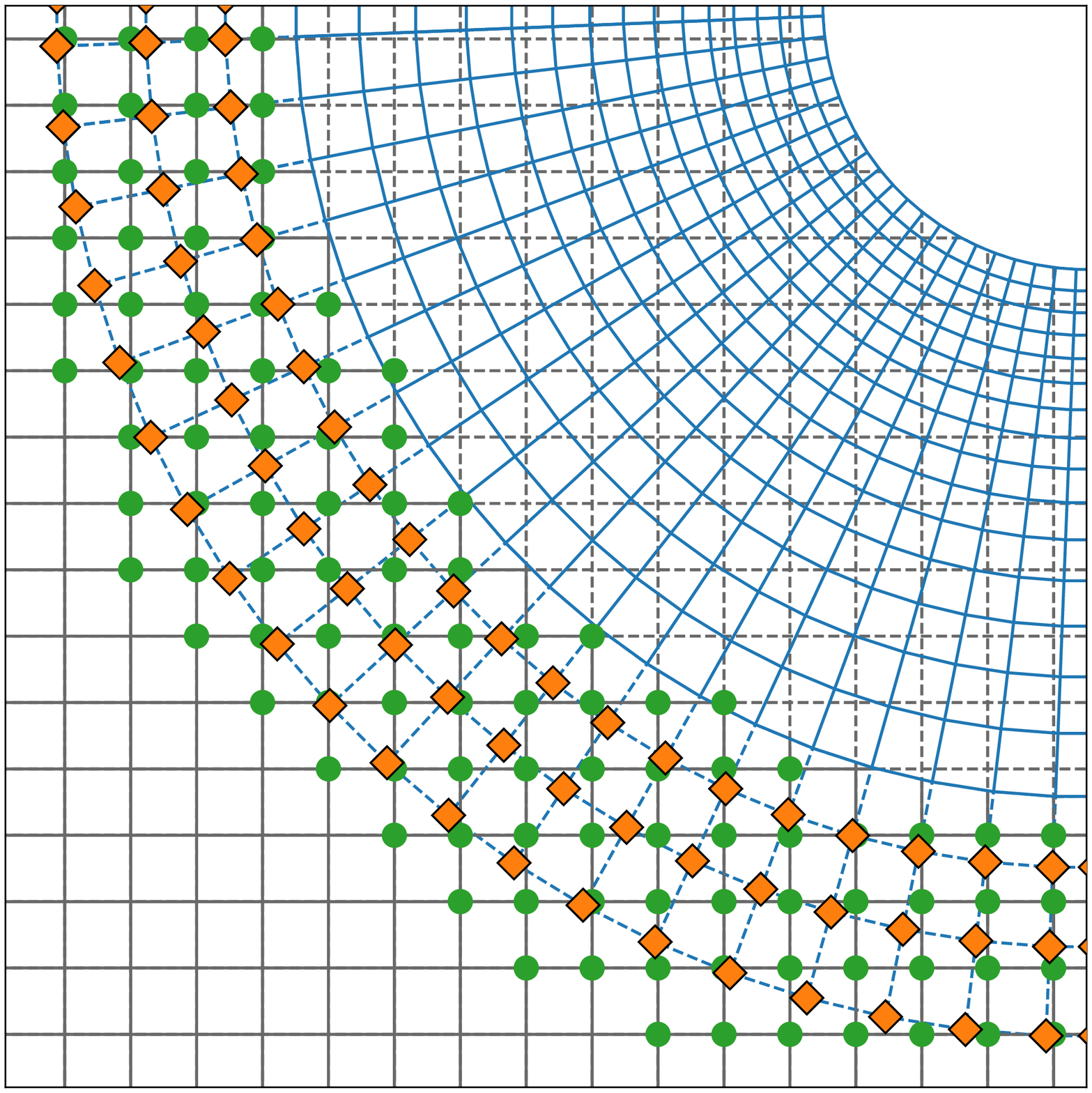}
    \caption{}
    \label{subfig:cart_to_curv}
  \end{subfigure}
  \begin{subfigure}[a]{0.49\textwidth}
    \centering
    \includegraphics[width=0.99\linewidth]{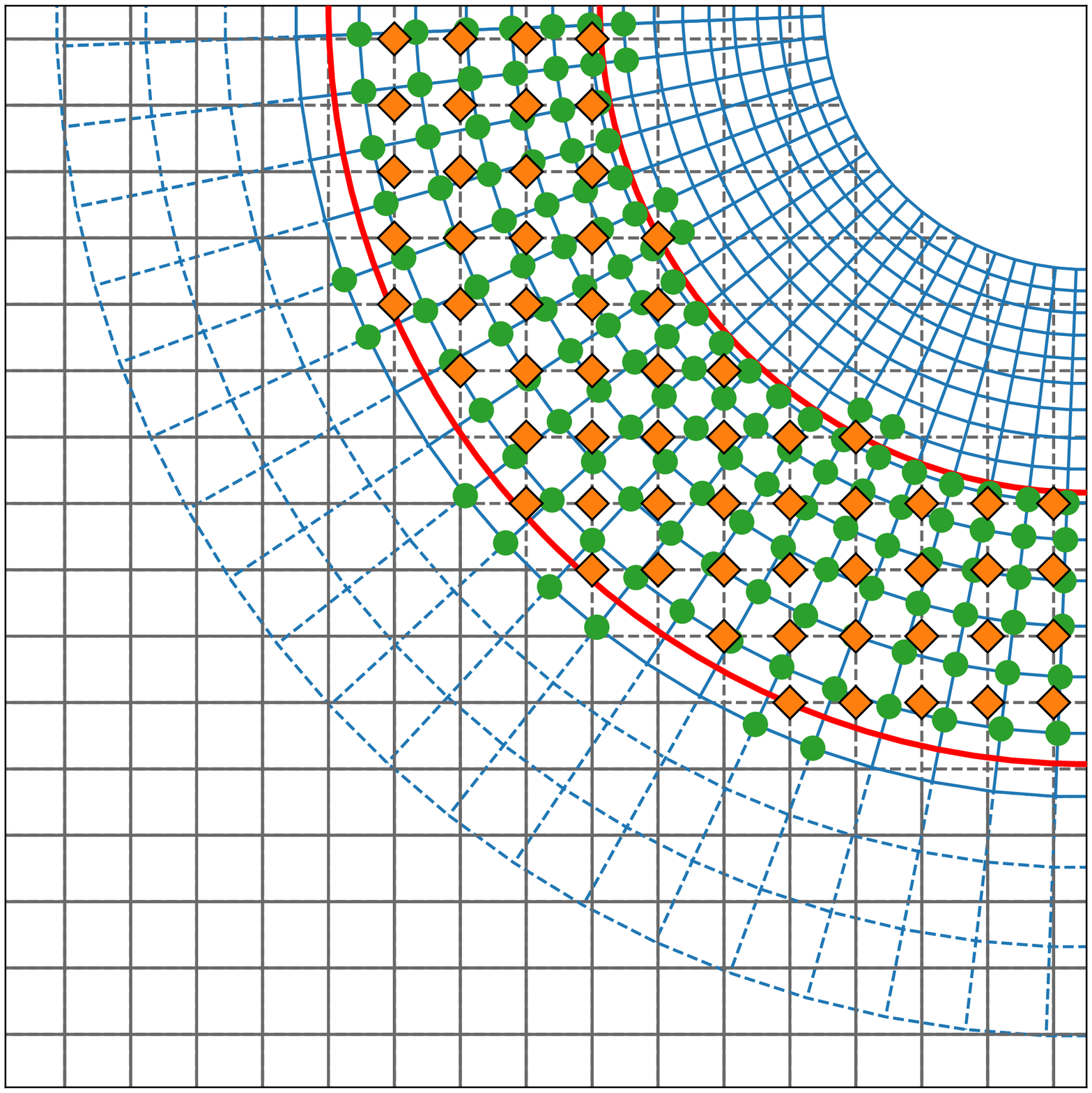}
    \caption{}
    \label{subfig:curv_to_cart}
  \end{subfigure}
  \caption{Overset grid method: (a) Cylinder grid on top of background grid (fringe-points of cylinder grid and background grid points within cylinder grid radius not shown). (b) Communication between grids, interpolation from Cartesian donor-points to cylindrical fringe-points. (c) Communication between grids, interpolation from cylindrical donor-points to Cartesian fringe-points. Four donor-points (\tikzcircle[mygreen, fill=mygreen]{2.5pt}) surround each fringe-point (\sqdiamond) in bi-linear interpolation. Dashed lines used where variables are not computed by finite-differences (fringe-points and hole-points).}
  \label{fig:overset_grid}
\end{figure}

To resolve the flow domain \rev{using} an overset grid method, a cylindrical coordinate grid \rev{was} body-fitted to the cylinder, and a uniform Cartesian grid \rev{was} used as the background grid (see Fig.~\ref{subfig:ogrid}). \rev{The cylindrical grid was streched} in the radial direction. In the region where fluid data is communicated between \rev{grids, it is beneficial that the grids have similar spacing. Grid} stretching enables similar grid spacing in the interpolation region and a much finer grid near the cylinder surface.
	
The compressibility of the flow lead to a strict stability limit for the Runge-Kutta method, imposing a very small time step in the simulations. \rev{Because the overset grids is} effectively solving two different flow problems, coupled only by the communication between the grids, \rev{there us} flexibility in the choice of time step. Choosing a time step on the background grid \rev{that is} small enough to guarantee stability for the Cartesian grid spacing, and \rev{choosing} a smaller time step on the cylindrical grid reduce\rev{s} the overall computational cost significantly. The cylindrical grid time step must be a multiple of the background grid time step to ensure that the computations on each grid are synchronized. \rev{An implicit solver may be beneficial}, if the grid spacing near the cylinder is several orders of magnitude smaller than that of the background grid, \rev{but this is} beyond the scope of this study.
	
The communication between the grids in the overset grid simulation \rev{was completed} in two stages for each time step of the background grid. At each stage of the communication, the required flow properties \rev{were} interpolated from donor-points to fringe-points. Each grid \rev{requires} a zone of fringe-points at least three points deep, such that the seven point central difference stencil \rev{could} be used without any special handling \rev{of} points adjacent to the fringe-points. For a curvilinear grid, the fringe-points \rev{were} simply the three outer points at each radial grid line (see Fig.~\ref{subfig:cart_to_curv}). For the Cartesian grid, the fringe-points \rev{must} be identified, typically during pre-processing, \rev{in order} to include all grid points within a fixed area in the region covered by both the Cartesian and the cylindrical grid. This is set by an inner and outer radius defining the interpolation region, see red lines enclosing fringe-points on the Cartesian grid in Fig.~\ref{subfig:curv_to_cart}. Cartesian grid points that are closer to the solid than the inner radius of the fringe-point zone (or inside the solid), are hole-points. The hole-points are not used in the computations.

In the overset grid method implemented in the Pencil Code, there is no overlap between the two interpolation regions of Figs.~\ref{subfig:cart_to_curv} and \ref{subfig:curv_to_cart}. That is, no fringe-points are used as donor points. Hence, the interpolation is explicit, not implicit \cite{Chesshire1990}. Note that if the bluff body enclosed by the body-fitted grid \rev{was} moving, the cost of inter-grid communication \rev{would be} significantly increased due to \rev{the} cost related to identifying new fringe and donor-points on the background grid at each new position of the bluff body. 
	
At present, two types of interpolation are implemented for overset grid communication in the Pencil Code: bi-linear Lagrangian interpolation and bi-quadratic spline interpolation. Both methods have the advantage of avoiding oscillations in the interpolation interval, \rev{which is} a common problem for high-order interpolation. The Lagrangian interpolation is a second-order accurate scheme, while the spline interpolation is third-order accurate. The illustration of donor-points and fringe-points in Figs.~\ref{subfig:cart_to_curv} and \ref{subfig:curv_to_cart} is for Lagrangian interpolation, where each fringe-point on one grid is interpolated from the $ 2 \times 2 $ surrounding donor-points \rev{of} the other grid. For spline interpolation, a zone of the $ 3 \times 3 $ closest grid points are used as donor-points for interpolation of each fringe-point. Note that the interpolation is bi-linear or bi-quadratic in both two- and three-dimensions. This is due to the Cartesian and cylindrical grid having a shared $ z $-plane, hence no interpolation is \rev{required} in the $ z $-direction.

At the solid-fluid interface, summation-by-parts finite-difference operators are used to enhance stability for unsteady flow \rev{simulations} (an unsteady wake develops for $ Re>47 $). \rev{These operators} are third-order \rev{accurate} for the sixth-order centered finite-difference method. Details on these operators can be found in \cite{Strand1994} (first derivatives) and \cite{Mattsson2004} (second derivatives).
	
The centered finite-difference schemes are non-dissipative\rev{, which can be detrimental} due to the potential growth of high-frequency modes, leading to numerical instability. To some extent, the summation-by-parts boundary conditions suppress such instabilities, but are not sufficient to suppress all oscillations in the solution on the curvilinear stretched grid. In particular, such oscillations are prominent in the density field. The detrimental effect of the high-frequency modes increase\rev{s} as the grid spacing decreases, \rev{which} may lead to diverging solutions as the grid is refined. To suppress the high-frequency modes, a high-order low-pass filter is used on the curvilinear \rev{portion} of the overset grid. The filter is a 10th order Pad\'{e} filter, with boundary stencils of 8th and 6th order. On the interior of the domain, the filter is given by: 
\begin{equation}
	\alpha_f \hat{\phi}_{i-1} + \hat{\phi}_{i} + \alpha_f \hat{\phi}_{i+1} =
	\sum_{n=0}^{N} \frac{\alpha_n}{2} ({\phi}_{i+n} + {\phi}_{i-n}) \, ,
\end{equation}
where $ \hat{\phi}_k $ and $ {\phi}_k $ are components $ k $ of the filtered and unfiltered solution vectors, respectively, $ \alpha_f $ is a free parameter ($\left| \alpha_f \right| \leq 0.5$) and $ \alpha_n $ are fixed parameters dependent only on $ \alpha_f $ \cite{Visbal1999}. Boundary stencils can be found in Gaitonde and Visbal \cite{Gaitonde2000}. The Pad\'{e} filter is implicit, and requires \rev{the solution of} a tri-diagonal linear system at grid point\rev{s} in the radial direction, and a cyclic tri-diagonal system at every grid point in the direction tangential to the surface. The free parameter $ a_f$ \rev{was} set to 0.1, \rev{where} filtering the solution once per Cartesian time step \rev{was} found \rev{to be} sufficient \rev{for} a stable and accurate solution.
	
\section{Performance}
\label{sec:benchmarking}
\subsection{Assessment of accuracy}
The spatial accuracy of the overset grid method \rev{was} examined by simulating a steady flow past a circular cylinder at \rev{a} Reynolds number \rev{of} 20. A domain of size $L_x \times L_y = 10D \times 10D $ \rev{was} used. The diameter of the curvilinear, body-fitted grid (henceforth called the cylinder grid) \rev{was} three times the cylinder diameter.
	
An indicative measure of the accuracy of the method can be found by computing solutions on several grid refinement levels, and using the finest grid as the ``correct solution" when computing two-norm errors. The grids used in this accuracy assessment are listed in Tab.~\ref{tab:accuracy_grids}. An odd number of grid points \rev{was} used in the directions that \rev{were} not periodic, \rev{in order} to \rev{have} grid points that are aligned at each refinement level. A fixed (dimensionless) time step $\Delta t = 0.25 \times 10^{-5} $ \rev{was} used for the Cartesian grid computations at all refinement levels. The small time step ensure\rev{d} that there \rev{was} no violation of diffusive or advective time step restrictions on any of the grids. These restrictions are $ \Delta \tau \leq C_{\nu} \Delta \chi_{min}^2 / \nu $ and $ \Delta \tau \leq C_u \Delta \chi_{min} / \left( \left| \vect{u} \right| + c_s\right)$, respectively, where $ \Delta \tau $ is the dimensional time step, $ \Delta \chi_{min} $ \rev{is} the smallest grid spacing in any direction, and $ C_\nu $ and $ C_u $ are the diffusive and advective Courant numbers, respectively. 

\begin{table}[t]
  \centering
  \caption{Grid refinement levels used in the assessment of accuracy
    of the overset grid method.}
  \label{tab:accuracy_grids}
  \begin{tabular}{c c c}
    \hline
    Refinement & Cylinder grid & Cartesian grid\Tstrut  \\
    level 	 &	$ N_r \times N_\theta $ & $ N_x \times N_y $\Bstrut \\
    \hline
    0	 & $ 17 \times 80 $   &	$ 80  \times 81 $\Tstrut \\	
    1	 & $ 33 \times 160 $  & $ 160 \times 161 $ \\
    2 	 & $ 65 \times 320 $  & $ 320 \times 321 $ \\
    3	 & $ 129 \times 640 $ & $ 640 \times 641 $\Bstrut \\
    \hline
  \end{tabular}
\end{table} 

Hyperbolic sine functions \rev{were} used for the stretching in the radial direction. The grid stretching \rev{parameters were} set such that the ratio between the grid spacing normal and tangential to the surface \rev{was} approximately one, both in the vicinity of the solid surface and in the interpolation region in the outer part of the cylinder grid. Furthermore, the number of grid points \rev{in} the Cartesian and cylindrical grids \rev{were} chosen \rev{in order} to have similar grid spacings in the region of inter-grid interpolation. The resulting local time step on the cylindrical grid \rev{was} $ \Delta t_{c} = 0.2 \Delta t $.

\begin{figure}[t]
  \centering
  \includegraphics[width=0.8\linewidth]{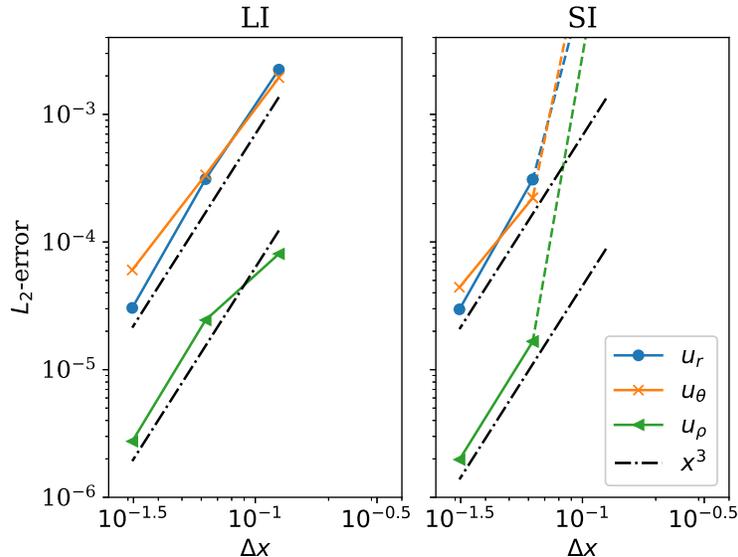}
  \caption{$ L_2 $-error norms of $ u_r $, $ u_\theta $ and $ \rho $
    at varying refinement levels at the grid point closest to the
    cylinder surface (for the coarsest grid). Results are for the computations with bi-linear Lagrangian interpolation \rev{(LI)} and bi-quadratic spline interpolation \rev{(SI)}, \rev{with} $ \Delta x $ \rev{(non-dimensional)} grid spacing on the Cartesian grid.}
  \label{fig:accuracy_grids}
\end{figure}

The main \rev{objective} of the method, is to compute a very accurate boundary layer around the cylinder. This is crucial for the application to particle impaction simulations in Section \ref{sec:particles} and in future studies. The $ L_2 $-error norms of flow variables  \rev{are therefore considered} along strips tangential to the cylinder surface as close as possible to the surface, \rev{in order} to get an indication of the accuracy of the scheme in the boundary layer. Figure \ref{fig:accuracy_grids} depicts $ L_2 $-error norms of the density and the normal and tangential velocity components (with respect to the cylinder surface), computed with the two different interpolation methods. The norms \rev{were} computed along a strip around the cylinder, at the grid point closest the cylinder for \rev{the} refinement level 0 (this corresponds to the 2nd point from the cylinder for refinement level 1, 4th for level 2, etc.). \rev{Note that the computations with spline interpolation did not fully converge to a stable solution at the coarsest grid level, as indicated by the dashed lines between the first refinement results in Fig.~\ref{fig:accuracy_grids}.}

For both interpolation methods computation of the density \rev{was} \rev{third-order}, the radial velocity component \rev{was} between third- and fourth-order \rev{and the tangiantal velocity component was between second- and third-order} accurate, at the grid point closest to the surface on the coarsest grid. \rev{The results suggest that the difference in accuracy between the interpolation methods is negligible in the immediate vicinity of the cylinder.}   


\begin{figure}[t]
	\centering
	\includegraphics[width=0.8\linewidth]{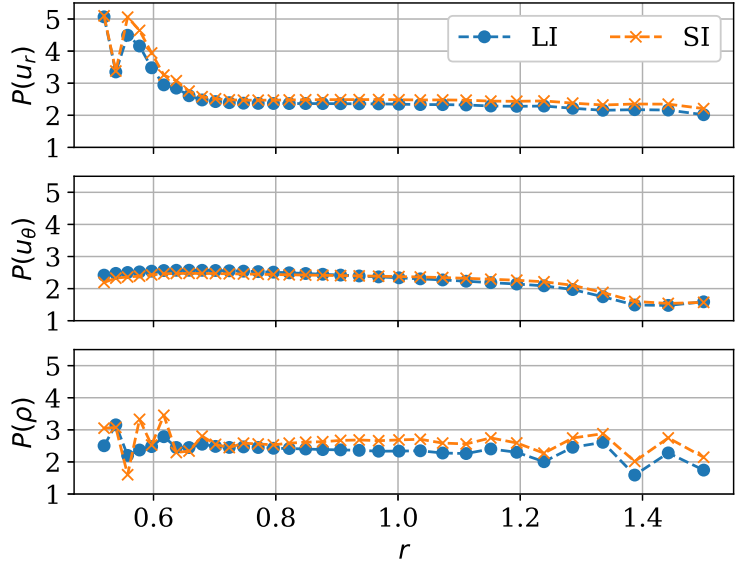}
	\caption{Formal order of accuracy of velocity components computed along strips tangential to the cylinder surface at non-dimensional radial position $ r $, for upper refinement levels for flow with $ Re=20 $ with Lagrangian interpolation (upper) and spline interpolation (lower). }
	\label{fig:accuracy_grids2}
\end{figure}

For a more detailed picture of the formal order of accuracy of the overset grid method, consider Fig.~\ref{fig:accuracy_grids2}. Th\rev{is} figure depicts the formal order of accuracy $P$, of the \rev{density and} velocity components, computed along strips at increasing distance from the cylinder boundary (cylinder boundary at $ r_c=0.5 $). \rev{The computations are} based on the assumption that the $ L_2 $-error norm on a grid with grid spacing $ \Delta x $ can be expressed as $ L_2(\Delta x) \sim \Delta x^p  $, \rev{such that} the order of accuracy $ P $ can be computed by:
\begin{equation}
	P = \frac{\log(L_2(\Delta x)/L_2(\Delta x / 2))}{\log2} \; .
	\label{eq:p}
\end{equation}
In principle, the spline interpolation scheme is third-order accurate while the Lagrangian interpolation a second-order accurate method. The effect of using the different methods of interpolation can be seen in Fig.~\ref{fig:accuracy_grids2}.
\rev{Some effect of the interpolation is seen, when considering the entire flow domain considered by the cylindrical grid. The difference is, however, much smaller than the difference in accuracy between the two interpolation schemes.} 
\rev{The difference in order of accuracy of the radial velocity computations is 0.02--0.56, for which spline interpolation yielded the highest order (median $P = 2.49 $ with spline interpolation, $ P = 2.42 $ with Lagrangian interpolation). A similar difference can be seen for the density. For the tangential velocity, on the other hand, there is no obvious best method. }
In the vicinity of the cylinder surface the difference between the Lagrangian and spline interpolation \rev{is negligible. Notice that the radial velocity component was computed with a} very high order of accuracy, $ P\approx5 $ \rev{in this region}. This is significantly more accurate than the more conservative suggestion of radial velocity accuracy between third- and fourth-order, \rev{seen} in Fig.~\ref{fig:accuracy_grids} (the results in Fig.~\ref{fig:accuracy_grids} correspond to the second point from the left in Fig.~\ref{fig:accuracy_grids2}).

\rev{The consideration of formal order of accuracy shows that the overset grid method is a high-order method ($P>2$). No obvious distinction among the two interpolation schemes was found by this analysis, although the spline interpolation appeared to have a marginally higher accuracy for the density and radial velocity component.}

\subsection{Unsteady flow {\label{subsec:transient}}}
The $ L_2 $-error norms are suggestive of the formal accuracy of the numerical method, but do not reveal the in-use accuracy of the method for simulations in the unsteady flow regime. \rev{This must be determined}, before arriving at a full-blown simulation of a particle-laden flow interacting with a cylinder in this flow regime.
 
A grid refinement study \rev{was} performed for $ Re=100 $, where unsteady vortex shedding developed in the cylinder wake. A domain with $ L_x \times L_y = 10D \times 20D $ \rev{was} used, with the cylinder in the center of the domain. The resulting mean drag coefficient ($ {C_D} $), root-mean-square lift coefficient ($ {C_L'} $) and Strouhal number ($ Str $) \rev{were} computed.  The drag and lift coefficients \rev{were} computed \rev{using} the pressure and shear forces on the cylinder, $\vect{F_p}$ and $\vect{F_s} $, respectively, \rev{as} given by:
\begin{align}
  \vect{F_p} &=-\int{p\big|_{r_c} \vect{dA}}  \approx 
  -\vect{\hat{r}} h  r_c\Delta \theta \sum_{i=1}^{N_\theta} p(r_c,\theta_i)  \; , \\
  \vect{F_s} &=\int\vect{\sigma}\big|_{r_c}dA  \approx 
  \vect{\hat{\theta}} \nu h r_c\Delta \theta \sum_{i=1}^{N_\theta}  \rho(r_c,\theta_i) \left.\diff{u}{r}\right|_{(r_c,\theta_i)}  \; ,
\end{align}
where $ h $ is the height of the cylinder and $ \vect{\sigma} $ is the shear stress. With flow in the y-direction, the drag and lift forces, $ F_D $ and $ F_L $, \rev{were} found by taking the sum of the pressure and shear forces in $ y $- and $ x $-direction, respectively. These forces can be used to \rev{calculate} the drag and lift coefficients \rev{as follows}:
\begin{align}
	C_D &= \frac{F_D}{\frac{1}{2} \rho_0 U_0^2 A} \; , \\
	C_L &= \frac{F_L}{\frac{1}{2} \rho_0 U_0^2 A} \; ,
\end{align}
where $ \rho_0 $ and $ U_0 $ are free-stream values of the density and velocity, respectively, and $ A = 2h r_c$ is the projected frontal area of the cylinder. The Strouhal number is simply the shedding frequency, non-dimensionalized by the free-stream velocity and cylinder diameter. 
	
\rev{A} grid refinement study of the unsteady flow \rev{was} performed with both Lagrangian and spline interpolation on two differently sized overset grids. One ha\rev{d} a cylindrical grid with diameter $ 3D $ \rev{(}the same size that was used in the assessment of accuracy for the $ Re=20 $\rev{)}. The other ha\rev{d} a size $ 5D $. Hence, there \rev{was} a factor two difference in the radial length ($ L_r = r_{cg}-r_c $, where $ r_{cg} $ is the outer cylinder grid radius) of the two cylindrical grids. At each refinement level, the smallest spacing in the radial direction \rev{was} the same for the two different overset grids, and the stretching \rev{properties were} the same as that in the $ Re=20 $ flow simulations (approximately quadratic cells in the vicinity of the surface and the interpolation region, and approximately equal grid spacing on the Cartesian and curvilinear grid \rev{in} the interpolation region). Hence, the outer grid spacing on the larger cylindrical grid \rev{was} larger than the outer grid spacing of the smaller cylindrical grid. Thus, a coarser Cartesian grid \rev{could} used for the overset grid with the larger cylinder grid. This, in turn, allow\rev{ed} for a larger time step on the background grid, but requir\rev{ed} more sub-cycles on the cylindrical grid for each Cartesian time step. Details on the grids used in this refinement study are listed in Tab.~\ref{tab:Re100cases}.

\begin{table}[t]
	\centering
	\caption{Grid refinement levels used in the grid refinement study for flow past a cylinder at $ Re=100 $ with two different\rev{ly} sized cylindrical grids. Grid spacing $ \Delta r $ non-dimensional\rev{ized} by the cylinder diameter.}
	\label{tab:Re100cases}
	\begin{tabular}{c c | c c | c c}
		\hline
		Refinement 	&	 $ \Delta r_{min} $ 	& \multicolumn{2}{c|}{$ r_{cg} = 3 r_c $}
		& \multicolumn{2}{c}{$ r_{cg} = 5 r_c $}\Tstrut  \\
		level 	 	&	$  \times 10^{-2} $ &$ N_r \times N_\theta $ & $ N_x \times N_y $&$ N_r \times N_\theta $ & $ N_x \times N_y $\Bstrut \\
		\hline
    0	 & $4.1  $ & $ 16 \times 80   $ & $ 80  \times 160  $ & $24\times80  $ & $50 \times100$\Tstrut \\	
		1	 & $2.7  $ & $ 24 \times 120  $ & $ 120 \times 240  $ & $36\times120 $ & $76 \times152$ \\
		2	 & $2.0  $ & $ 32 \times 160  $ & $ 160 \times 320  $ & $48\times160 $ & $100\times200$ \\
		3	 & $1.6  $ & $ 40 \times 200  $ & $ 200 \times 400  $ & $60\times200 $ & $128\times256$ \\
		4	 & $1.3  $ & $ 48 \times 240  $ & $ 240 \times 480  $ & $72\times240 $ & $150\times300$ \\
		5  & $0.97 $ & $ 64 \times 320  $ & $ 320 \times 640  $ & $96\times320 $ & $200\times400$ \\
		6  & $0.77 $ & $ 80 \times 400  $ & $ 400 \times 800  $ & $120\times400$ & $256\times512$ \\
		7  & $0.64 $ & $ 96 \times 480  $ & $ 480 \times 960  $ & $144\times480$ & $306\times612$ \\
    8	 & $0.48 $ & $ 128 \times 640 $ & $ 640 \times 1280 $ 	& $192\times640$ & $408\times816$\Bstrut \\
		\hline
	\end{tabular}
\end{table}

\begin{figure}[t]
	\centering
	\includegraphics[width=0.99\linewidth]{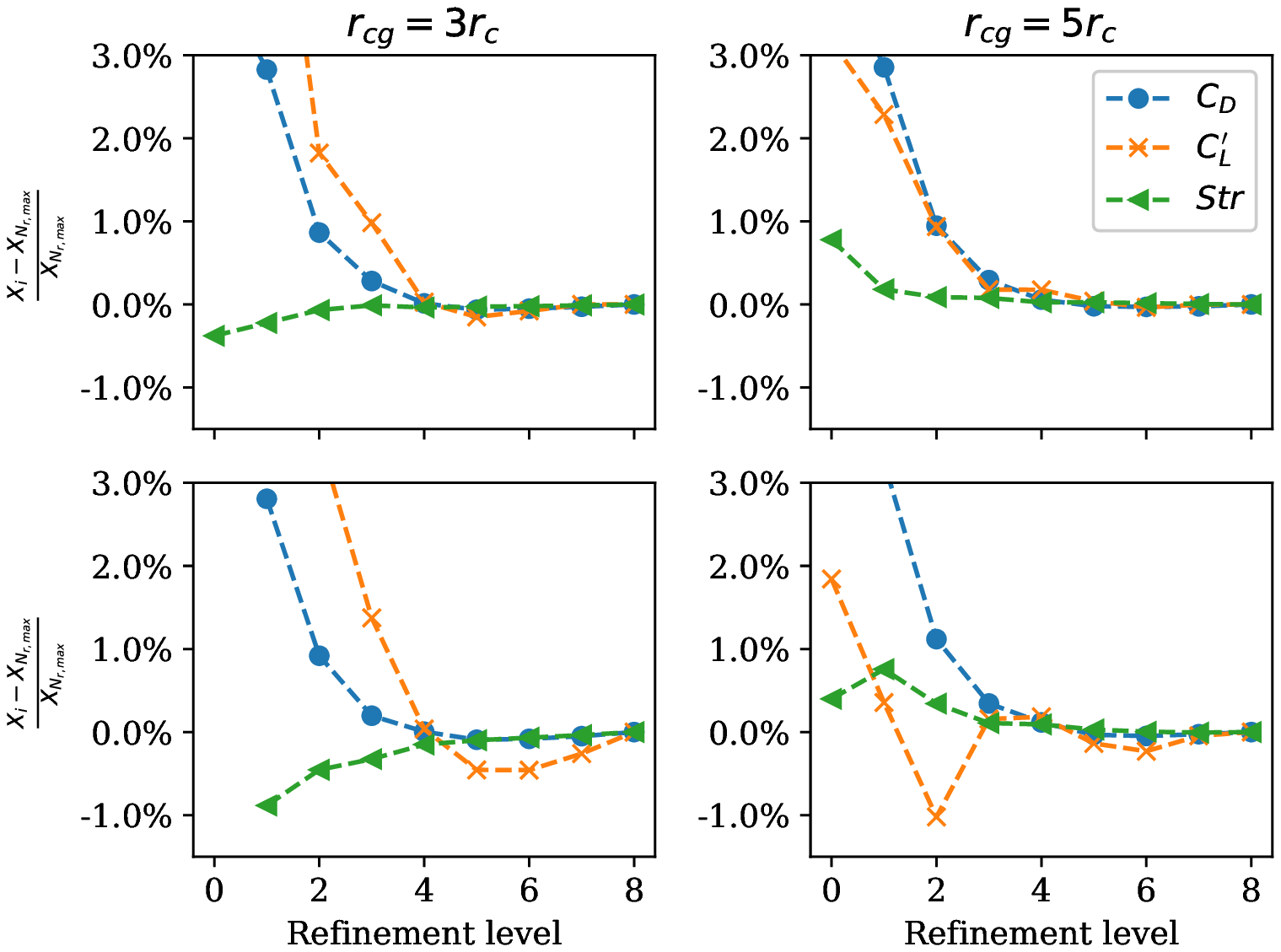}
	\caption{Normalized values for mean drag coefficient $(C_D) $, rms-lift coefficient $(C_L')$ and Strouhal number $(Str)$ for $ Re=100 $ computed at different refinement levels (see Tab.~\ref{tab:Re100cases}) for overset grids with two sizes of radii \rev{for} the body-fitted cylindrical grid, $ r_{cg} $. Results are given for computations with Lagrangian interpolation (upper) and spline interpolation (lower).}
	\label{fig:gridref}
\end{figure}

Results for the grid refinement at $ Re=100 $ can be seen in Fig.~\ref{fig:gridref} and Tab.~\ref{tab:gridref_3r5r}. In Fig.~\ref{fig:gridref}, the dimensionless drag and lift coefficients, and the Strouhal number have been normalized by the result computed at the finest grid. Hence, the plots depict the relative deviation from the result at grid refinement level eight from Tab.~\ref{tab:Re100cases}. The values of the coefficients computed at this refinement level, for each of the four cases, are given in Tab.~\ref{tab:gridref_3r5r}.

\begin{table}[t]
	\centering
	\caption{Mean drag coefficient $(C_D) $, rms-lift coefficient $(C_L')$ and Strouhal number $(Str)$ for $ Re=100 $ computed at a domain $ L_x \times L_y = 10D \times 20D$ with two different overset grids. The resolution is given by the finest refinement levels in Tab.~\ref{tab:Re100cases}, and both Lagrangian (LI) and spline interpolation (SI) cases are considered. }
	\label{tab:gridref_3r5r}
	\begin{tabular}{c | c c | c c}
		\hline
		 	&	  \multicolumn{2}{c|}{$ r_{cg} = 3 r_c $}
		& \multicolumn{2}{c}{$ r_{cg} = 5 r_c $}\Tstrut  \\
		Coefficient&  LI & SI & LI & SI\Bstrut \\
		\hline
		$ C_D $ & 1.46\rev{0} & 1.45\rev{8} & 1.4\rev{61} & 1.4\rev{61}\Tstrut \\
		$ C_L'$ & 0.25\rev{09} & 0.24\rev{50} & 0.2\rev{527} & 0.2\rev{522} \\
		$ Str $ & 0.17\rev{63} & 0.17\rev{62} & 0.17\rev{63} & 0.17\rev{63}\Bstrut \\
		\hline
	\end{tabular}
\end{table}

\rev{T}he dimensionless numbers converge\rev{d} quite rapidly for all of the \rev{tested} cases. \rev{The best performance for grid independency was achieved with Lagrangian interpolation. Yet, even the poorest result, the computation of the rms-lift coefficient at the smaller of the two cylindrical grids with spline interpolation deviated less than 0.5\% from the finest grid result, for grid refinement level $ \geq 4 $.} \rev{Some deviation is also seen in the lift coefficient of results computed on the large cylindrical grid with spline interpolation (less than 0.24\% for refinement level $ \geq 4 $). For the cases where Lagrangian interpolation was used for inter-grid communication,} the deviation from results at refinement levels four to seven from the finest grid result is less than $ 0.1\rev{5}\% $ for all coefficients. (if only drag and Strouhal number are considered, the deviation is less than $ 0.06\rev{4}\% $ for these cases). 

\rev{The difference between results obtained with quadratic spline and linear Lagrangian interpolation was particularly clear for the smaller cylinder grid. With a larger grid, it is not surprising that the effects from interpolation were reduced. Nevertheless, the best results on the larger grid were also achieved with Lagrangian interpolation. For the steady flow simulations the spline interpolation yielded results with \rev{somewhat higher} order of accuracy. The sub-par performance of this interpolation for unsteady simulations may be due to the overshoots in this non-linear interpolation, or perhaps, a larger mass loss during interpolation. No further speculation is conducted here}, but note that the Lagrangian interpolation outperformed the spline interpolation \rev{for simulations of unsteady flow.}

By considering the grid independent solutions in Tab.~\ref{tab:gridref_3r5r}, used to normalize the grid refinement results, two particular factors \rev{could be noted}. Firstly, by comparing the results \rev{for} the two different interpolation schemes \rev{on} the domain where the cylindrical grid has $ r_{cg}=5r_c $, \rev{it is evident that} the computed drag, lift and Strouhal number \rev{were} independent of \rev{the} inter-grid communication. This in contrast to the $ r_{cg}=3r_c $ results, but in accordance with \rev{an} intuitive understanding of the problem: the farther away from the cylinder boundary the interpolation is \rev{performed}, the less it affects computation of quantities at the boundary. Note, however, that even though the drag and lift forces \rev{were} computed at the boundary, these coefficients \rev{were also} dependent on the flow upstream and downstream of the cylinder. The results therefore suggest that the flow surrounding the cylinder \rev{was negligibly impacted by} interpolation method \rev{selected} when the larger $ r_{cg} $ \rev{was} used for the cylinder grid.

\rev{B}y comparing the results for $ C_D $ and $ C_L' $ on the different\rev{ly} sized cylinder grids, computed with Lagrangian interpolation, the \rev{dependency on cylinder grid size was found to be small. There was a small difference in the computed lift coefficient (somewhat higher for the larger cylinder grid).} Although the results are grid independent, neither of the values are quantitatively accurate for the drag or lift of a cylinder in a cross flow at $ Re=100 $. \rev{The small difference in computed lift may be due to blockage effects \rev{or} interpolation errors propagating across periodic boundaries.}

To \rev{confirm} that the grid independent solutions yield\rev{ed} accurate flow predictions a simulation \rev{was also conducted} on a large domain, $ L_x \times L_y = 50D \times 50D $, for \rev{the two different grid sizes used above. Since Lagrangian interpolation had the best performance for the unsteady flow simulations, only this interpolation procedure was used.} The grid spacing corresponding to grid refinement level five in Tab.~\ref{tab:Re100cases} \rev{was} used on the large square domain. The computed flow quantities show\rev{ed} good agreement with previous studies performed on similar domain sizes (see Tab.~\ref{tab:Re100comparison}). \rev{Note that for the simulations on a large square domain there was a negligible difference between the results from the different overset grid simulations. Because a smaller domain was used in the next section (with width $ 6D $), the smallest cylinder grid (with  $r_{cg}= 3r_c $) was selected. Using a larger cylinder grid will reduce the total number of grid points in the simulations, which is a major advantage on large domains. On small flow domains, overset grids with interpolation regions very close to each side of the periodic boundary should not be used, as this may cause spurious interpolation errors to propagate across the boundary.} 

\begin{table} \centering
  \caption{Comparison with previous studies. The studies \rev{were} performed
    on domains with streamwise length $ 60 \leq L_x/D \leq 100 $ and
    spanwise length $ 40 \leq L_x/D \leq 100 $, and the present study
    has $ L_x= L_y = 50D $. Results from the present study are for domains covered by two different\rev{ly} sized overset grids, with inter-grid interpolation performed by bi-linear Lagrangian interpolation.
    The asterisk on some values of $ C_L' $ denote\rev{s} where only
    the amplitude of the lift was given. The asterisk mark a lift amplitude scaled by $ 1/\sqrt{2} $ to get the root-mean-square lift coefficient, a valid scaling for the sinusoidal-like lift coefficient (with mean value zero).}
  \label{tab:Re100comparison}
  \begin{tabular}{l l l l}
    \hline
     & $ C_D $ & $ C_L' $ & $ St  $\Tstrut \Bstrut \\
    \hline
    Li et al.~\cite{Li2009}	& 1.336  &   $\quad -$ 	    &  0.164\Tstrut    \\
    Posdziech \& Grundmann \cite{Posdziech2007}  & 1.350  & $0.234^{(*)}$  & 0.167 \\
    Pan	\cite{Pan2006} & 1.32 & $0.23^{(*)}$ & 0.16 \\
    Qu et. al.~\cite{Qu2013}  & 1.326 & 0.219 & 0.166 \\
    Present, $ r_{cg}=3r_c $  & 1.3\rev{46} & 0.235 & 0.16\rev{6} \\
	Present, $ r_{cg}=5r_c $  & 1.3\rev{46} & 0.23\rev{4} & 0.16\rev{6} \Bstrut \\
    \hline 
  \end{tabular}
\end{table}

\section{Particle deposition on a circular cylinder in a laminar cross flow {\label{sec:particles}}}
Direct numerical simulations with a large number of particles suspended in the flow have been performed to assess the performance of overset grids on a more complex and demanding simulation than the simple flow past a cylinder at low Reynolds numbers.
	
The particle deposition simulations are based on the study by Haugen \& Kragset \cite{Haugen2010}\rev{,} where particle-laden flow simulations were performed \rev{over} a range of Reynolds numbers on a moderate\rev{ly} sized flow domain ($ 6D \times 12D $). The analysis \rev{is} not repeated here, \rev{but} a brief introduction to the method used for particle representation and deposition \rev{will be} present\rev{ed}. The particle-laden flow simulations \rev{were} performed on a domain \rev{exactly} the same size as in Haugen \& Kragset\cite{Haugen2010}, with Reynolds number 100.

\subsection{Particle equations}
The particles are tracked using a Lagrangian formalism, where the particle velocity and position are described by:
\begin{align}
\label{eq:vp}
\frac{\mathrm{d}\vect{v}_p}{\mathrm{d}t} &= \frac{\vect{F}_ {D,p}}{m_p} \, , \\
\frac{\mathrm{d}\vect{x}_p}{\mathrm{d}t} &= \vect{v}_p \, ,
\end{align}
where $ \vect{v}_p $, $ \vect{x}_p $ and  $ m_p $ are the velocity, center of mass position and mass of the particle, respectively. The force $ \vect{F}_ {D,p} $ acting upon a spherical particle is the drag force:
\begin{equation}
\vect{F}_ {D,p} = \frac{1}{2C_c} \rho C_{D,p} A_p \left|\vect{u}-\vect{v}_p\right|\left(\vect{u}-\vect{v}_p\right) \, ,
\end{equation}
where $ A_p = \pi d_p^2/4$ is the cross sectional area of the particle and
\begin{equation}
C_c = 1 + \frac{2\lambda}{d_p} \left(1.257 +  0.4e^{\left(-1.1d_p/2\lambda\right)} \right) \, ,
\end{equation}
is the Stokes-Cunningham factor (with parameters set for air) for a particle with diameter $ d_p $. The mean free path $ \lambda = 67 \, \mathrm{nm}$ accounts for the fact that for very small particles, the surrounding medium can no longer be regarded as a continuum but rather \rev{as} distinct particles. \rev{T}he particle drag coefficient is given by:
\begin{align}
C_{D,p} =
\frac{24}{Re_p} \left(1 + 0.15 Re_p^{0.687}\right) \, ,
\end{align}
for particle Reynolds number $ Re_p = d_p \left| \vect{v}_p - \vect{u} \right|/\nu \lesssim 1000 $. With this, the particle drag force can be re-written as
\begin{equation}
\vect{F}_ {D,p} = \frac{m_p}{\tau_p} \left(\vect{u}-\vect{v}_p\right) \, ,
\end{equation}
where
\begin{equation}
\label{eq:tau_p}
\tau_p = 
\frac{S d_p^2 C_c}{18 \nu(1 + f_c)}
\end{equation}
is the particle response time, with $ f_c = 0.15 Re_p^{0.687} $ and $ S = \rho_p / \rho $. Note that this \rev{is} Stokes drag in the limit $ C_c = 1 $ and $ f_c = 0 $. Using the convention of \cite{Haugen2010}, the Stokes number ($St = \tau_p / \tau_f$) is defined with a fluid time scale\rev{:}
\begin{equation}
\label{eq:tau_f}
\tau_f = \frac{D}{2 U_0} \, .
\end{equation}

The fluid velocity \rev{was} interpolated from surrounding grid points by bi-linear interpolation on the Cartesian grid and bi-quadratic interpolation on the curvilinear grid. The order of the interpolation is higher on the curvilinear grid as the velocity components (the radial, in particular) are close to quadratic near the cylinder surface. For three-dimensional simulations, linear interpolation is used for the velocity component along the z-direction (the cylinder's spanwise direction) on all grids.

For particles very close to the cylinder surface, special handling \rev{was} used to interpolate the radial component of the fluid velocity. \rev{V}ery close to the cylinder \rev{refers to within one grid point from} the surface, or alternatively, within the pre-calculated momentum thickness of the boundary layer. The special handling in use for particles at such positions \rev{was} a quadratic interpolation that guarantees no overshoots. Since all velocities are zero at the surface, this \rev{was} achieved by:
\begin{equation}
	\label{eq:u_r}
	u_{r,p} = u_{i,g} \left(\delta r_p/ \delta r_g\right)^2 \, ,
\end{equation}
where $ u_{r,p} $ and $ u_{r,g} $ are radial velocity components at the position of the particle and at the position of the interception between a surface normal and the first grid line \rev{away} from the surface, respectively. The distances $ \delta_p $ and $\delta_g$ are from the surface to the particle and to said grid line, respectively.

\begin{figure}[t]
	\centering
	\begin{subfigure}{0.32\textwidth}
		\centering
		\includegraphics[width=1.\linewidth]{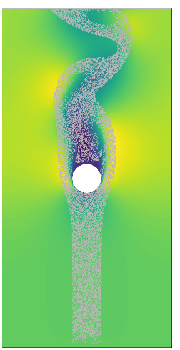}
		\caption{$ St=0.1 $}
		\label{fig:part_flow1}
	\end{subfigure}
	\begin{subfigure}{0.32\textwidth}
		\centering
		\includegraphics[width=1.\linewidth]{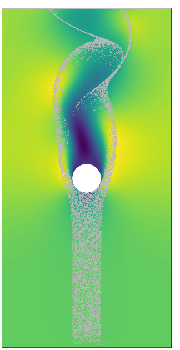}
		\caption{$ St=1.0 $}
		\label{fig:part_flow2}
	\end{subfigure}
	\begin{subfigure}{0.32\textwidth}
		\centering
		\includegraphics[width=1.\linewidth]{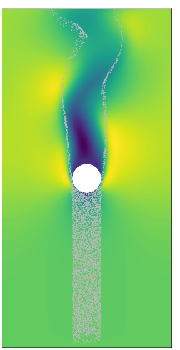}
		\caption{$ St=10 $}
		\label{fig:part_flow4}
	\end{subfigure}
	\caption{Particle-laden flow interacting with a circular cylinder at $ Re=100 $. An unhindered particle will cross the flow domain, from the inlet (bottom) to the outlet (top) in approximately two shedding periods at this Reynolds number. Contours of the streamwise velocity component make up the background.}
	\label{fig:part_flow}
\end{figure}
\subsection{Particle impaction {\label{subsec:imp_res}}}
After the flow developed \rev{in}to periodic vortex shedding, particles \rev{were} inserted continuously at the inlet. The particles \rev{were} inserted randomly, as a homogeneous distribution over a \rev{rectangular cross-section encompassing} particle trajectories that \rev{could} impact the cylinder. From here \rev{the particles were} convected downstream, and removed from the flow either by impacting the cylinder or by reaching the outlet (see Fig.~\ref{fig:part_flow}). An impaction \rev{was} registered (and the particle removed) if the distance between the cylinder surface and the particle's center of mass \rev{was} less than or equal to $ d_p/2 $. Every particle impaction simulation \rev{was} run until all particles \rev{were} removed from the flow. In total $ 1.1 \times 10^7 $ particles \rev{were} inserted, with Stokes numbers of 0.01--10, \rev{and} a progressive particle distribution with respect to particle radius.

\begin{figure}
  \centering
	  \includegraphics[width=0.7\linewidth]{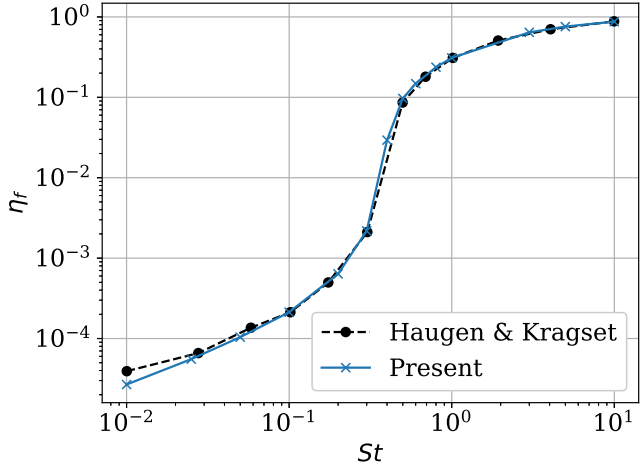}
  \caption{Front side impaction efficiency ($ \eta_f $) as a function
    of Stokes number ($ St $) for Reynolds number 100. Present results
    compared to a previous study by Haugen \& Kragset
    \cite{Haugen2010}}
  \label{fig:part_impact}
\end{figure}

The impaction efficiency ($ \eta = N_{imp}/N_{ins} $) can be split into front ($ \eta_f $) and back side impaction ($ \eta_b $). At the low Reynolds number flow in this study, backside impaction \rev{rarely occurred so} and front side impaction \rev{was the focus}. Figure \ref{fig:part_impact} depicts the particle front side impaction, compared to \rev{literature} results. The results \rev{were} computed with grid spacing defined as refinement level four in Tab.~\ref{tab:Re100cases}, for the $ r_{cg}=3r_c $ case with Lagrangian interpolation. With the $ L_x \times L_y = 6D \times 12D$ domain a grid $(N_r \times N_\theta) + (N_x \times N_y) = (48 \times 240) + (144 \times 288) $ \rev{was used}. The results from Haugen \& Kragset \cite{Haugen2010} were computed on an equidistant grid with $ 512 \times 1024 $ grid points, using an immersed boundary method to resolve the cylinder surface.
	
The particle impaction results from \rev{this} study agree very well with the results from \rev{the} literature, \rev{even} though the results \rev{of} the present study \rev{were} computed on \rev{a} grid with only 10.1\% as many grid points as used by Haugen \& Kragset \cite{Haugen2010}. An additional efficiency improvement was achieved\rev{,} due to us\rev{ing} a  time step \rev{that was} 3.5 \rev{times} larger. \rev{This was possible because of} the time step\rev{'}s proportionality to the grid spacing \rev{and the local time step restrictions}, though some extra work \rev{was necessary} at each time step (computation on two grids, communication of data, filtering on cylinder grid, etc.). Note that for very small particles, the time step can also be restricted by the particle time scale\rev{; t}hat is, the time step must be small enough to resolve the time-dependent particle equations. Particles are updated only at the Cartesian time step.

\subsection{Investigating the accuracy of the computed impaction efficiencies}
The coarseness of the grid used in the computation of particle impaction efficiencies allow \rev{for the assessment} of the assumptions that must be made in order to regard these impaction results as quantitatively accurate. The assumptions are, firstly, that blockage effects from the limited domain (with $ L_x \times L_y = 6D \times 12D$) have a negligible impact on the particle impaction. Secondly, \rev{it was assumed that} the coarsest resolution where grid independency of drag and lift coefficients was reached \rev{was sufficiently fine for the particle simulations}, i.e., that that the transport of the particles \rev{was} dependent on an accurate flow field only.

A critical assessment of these assumptions \rev{led to the expectation of} a higher impaction result for particles on a domain where the blockage effect is large, due to a squeezing of the flow field and, consequently, less particles being directed away from the cylinder. In particular, this is expected to affect particle\rev{s} that follow the flow to a large extent, i.e. particles with small Stokes numbers. Further, the flow velocities at particle positions are not only dependent on an accurately computed flow field, but also \rev{on} accurate interpolation. The latter aspect can be very sensitive to grid spacing, even if the flow is resolved accurately. Haugen \& Kragset \cite{Haugen2010} used linear interpolation to compute flow velocity at particle positions, except within the grid point closest to the surface, where an expression similar to that of Eq.~\ref{eq:u_r} was used. Linear interpolation of velocities that are proportional to $ -(\delta r)^2 $ (as the upstream flow field at the centerline through the cylinder is) will lead to a systematic over-estimation. Hence, an over-prediction in particle impaction can be expected from their results. What is important \rev{to determine} in this respect, is how large this possible over-prediction is, and for what particle sizes it occurs.

\begin{figure}
	\centering
	\includegraphics[width=1.\linewidth]{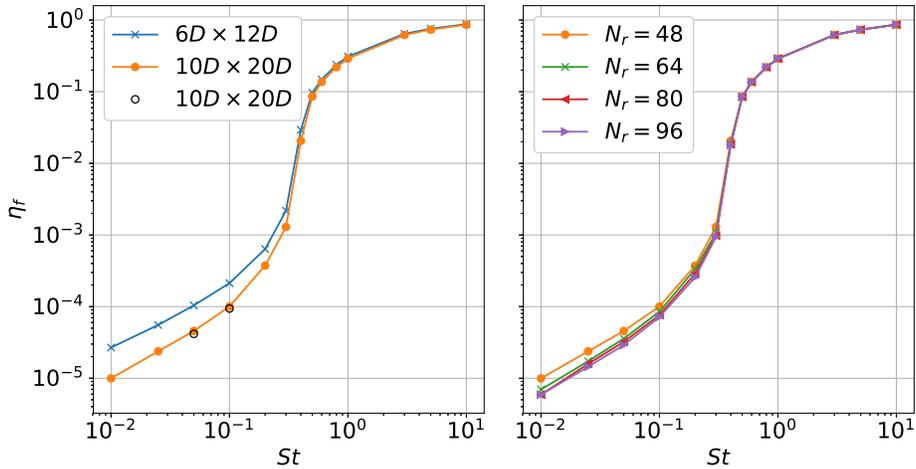}
	\caption{Front side impaction efficiency ($ \eta_f $) as a function of Stokes number ($ St $) for Reynolds number 100 for different domain sizes (left) and grid resolutions (right).}
	\label{fig:part_impact_comp}
\end{figure}

To investigate \rev{the} accuracy of the computed impaction efficiencies particle-laden flow simulations \rev{were conducted} at a larger domain size, $ L_x \times L_y = 10D \times 20D $, \rev{as used} in the grid independence study of Section \ref{subsec:transient}. \rev{For} this larger domain several refined grids \rev{were} used. These utilized refinement levels 4--7 in Tab.~\ref{tab:Re100cases}, with $ r_{cg} = 3r_c $. Thus, from 48 (coarsest) to 96 (finest) grid points \rev{were} used in the radial direction on the cylindrical grid, and the background grid \rev{was} refined accordingly. The number of inserted particles \rev{was} $ 1.1 \times 10^7 $, where $ 7 \times 10^6 $ \rev{were} particles with $ St\leq 0.1 $. The results are seen in Fig.~\ref{fig:part_impact_comp}.

Very few of the smallest particles deposit on the cylinder. To get enough particle impaction at the smallest Stokes numbers \rev{for} reliable statistics, particles with $ St\leq 0.1 $ \rev{were only inserted} over a region covering one tenth of the cylinder's projected are\rev{a}, \rev{at its} centerline. \rev{The inserted particle count was scaled correspondingly (multiplied by ten) during post-processing.} No small Stokes number particles inserted outside the insertion area \rev{would be} expected to hit the cylinder. To \rev{confirm} this, a simulation \rev{was conducted with particles} with $ St = 0.05 $ and 0.1, inserted over the whole projected cylinder area. \rev{T}he results are included as black circles (o) in Fig.~\ref{fig:part_impact_comp}. The \rev{difference of impaction efficiency among particles inserted by the two different methods was negligible.}

From Fig.~\ref{fig:part_impact_comp} it is clear that the blockage effect from the limited domain size \rev{had} a significant effect on the particle impaction efficiencies. For $ St \leq 0.5 $ this effect \rev{was} larger than 10\%, and increase\rev{d} as the Stokes number decreased. The largest difference in impaction efficiencies \rev{was} seen at $ St=0.01 $, where \rev{2.7 times} more impaction \rev{occurred} for the smallest domain size. The resolution played a smaller, but not insignificant, role in the impaction efficiencies. Increasing from the coarsest grid, with $ N_r = 48$, to $ N_r = 64 $, noticeably reduced the impaction efficiencies. The reduction \rev{was} more than 10\% for $ St \leq 0.3 $. A further refinement of the grid \rev{had} a small effect, \rev{which was} negligible for $ N_r \geq 80 $. Comparing the results from the larger domain with $ N_r=80 $ to \rev{those} by Haugen \& Kragset \cite{Haugen2010} suggests that Haugen \& Kragset found a qualitatively correct result, but have somewhat quantitatively over-predicted the particle impaction, in particular in the boundary interception region \rev{(where} $ St\lesssim 0.3 $). For the smallest Stokes number ($ St=0.01 $) the over-prediction is of approximately a factor 6.3. At $ St=0.1 $ this factor \rev{is 2.8}. \rev{The previously published results agree with the new results for $ St\geq0.5 $.}

\section{Concluding remarks \label{sec:conclusion}}

In this work, a high-order overset grid method \rev{has been presented}. The method uses high-order finite-difference discretization to solve the compressible Navier-Stokes equations on several grids, and communicates necessary flow data between the grids by linear or quadratic interpolation. Unique \rev{to} the overset grid implementation described here, is the use of local time step restriction and summation-by-parts finite-difference operators. The relaxed time stepping restriction on the coarser grid is very efficient for a weakly compressible flow, while the summation-by-parts operators enhance numerical stability together with the use of Pad\'{e} filtering. The purpose of developing the method \rev{was} to compute particle impaction on a cylinder in a cross flow, and for this purpose a body-fitted cylindrical grid is an appropriate choice to resolve the boundary layer around the cylinder with high accuracy. 

An investigation of the formal order of accuracy of the overset grid implementation revealed that high-order accuracy \rev{was} indeed reached. \rev{Flow variables where computed with median order $ P \approx 2.5 $, regardless of the use of} 
\rev{bi-linear interpolation or bi-quadratic interpolation for communication. Near the surface, the radial velocity component reached an accuracy of fifth-order.}
For unsteady flow, the method converge\rev{d} rapidly to grid independent solutions for the essential flow variables (drag, lift and Strouhal number).
\rev{For these computations, using bi-linear interpolation was beneficial, yielding the most rapid convergence to grid independent solutions as the grid was refined.}
Using a larger cylindrical grid, with \rev{a} radius five times as large as the cylinder radius, decreased the effect of the \rev{inter-grid} interpolation.

When applied to the problem of inertial particles impacting on a cylinder, impaction efficiencies of previously published results \rev{were} reproduced at a significantly reduced \rev{computational} cost. \rev{A} coarser background grid \rev{was utilized} to resolve the flow, which yielded both a much smaller number of grid point\rev{s} (90\% reduction in 2D) and the possibility \rev{of using} a larger time step.

A critical assessment of the particle impaction results revealed that the limited domain size had a significant impact on particle impaction, particularly for the smaller Stokes numbers. Further, although the flow was deemed grid independent, using a finer grid, and thus a more accurate interpolation of flow velocity, reduced the number of particles that hit the cylinder. The resulting impaction curves suggest\rev{ed} that particle impaction has been over-estimated in \rev{the} previous studies, in particular for very light particles where impaction occurs by boundary interception.

The overset grid method implementation in the Pencil Code is \rev{now} ready for three-dimensional simulations, and DNS studies of particle impactions on a cylinder \rev{with} Reynolds number for real-world application (\rev{a} factor 10-20 larger than the investigation here, for industrial boilers) is within reach. However, even with the highly accurate and efficient method presented here, increasing the Reynolds number and computing three-dimensional flow will be computationally costly. \rev{The magnitude of the} Reynolds numbers that can be considered will largely \rev{depend} on the Stokes numbers of the particles, and \rev{the} acceptable accuracy when particle impaction efficiencies are computed. If the focus is not just qualitative trends, but quantitatively accurate results, a careful assessment of grid independence \rev{is recommended} (not just \rev{for} flow variables, but \rev{for} the particle impaction itself), and \rev{great} care \rev{is required when selecting the} domain size and \rev{setting up the simulations.}

\section{Acknowledgements}
We would like to acknowledge \rev{the} valuable discussions on numerical methods and code development with our colleagues Prof. Bernhard M\"{u}ller and Dr. Ehsan Khalili.

The work was supported by the Research Council of Norway (Norges Forskningsr\aa d) under the FRINATEK Grant [grant number 231444]. Partial funding was provide by the GrateCFD project, which \rev{was} funded by: LOGE AB, Statkraft Varme AS, EGE Oslo, Vattenfall AB, Hitachi Zosen Inova AG and Returkraft AS together with the Research Council of Norway through the ENERGIX program [grant number 267957/E20]. Computational resources were provided by UNINETT Sigma2 AS [project numbers NN9405K and NN2649K]. 

\bibliography{./ref}
\end{document}